\documentclass[%
 reprint,
 amsmath,amssymb,
 aps,
pra,
]{revtex4-2}
\usepackage[utf8]{inputenc}
\usepackage{amsmath}
\usepackage{amsfonts}
\usepackage{braket}
\usepackage{comment}
\usepackage{graphicx}
\usepackage{physics}
\usepackage{xcolor}
\usepackage{multirow}
\usepackage{adjustbox}% bold math
\usepackage{hyperref}
\begin{document}
\preprint{APS/123-QED}
\title{Mueller matrix based characterization of Cervical tissue sections: A quantitative comparison of Polar and Differential decomposition methods}% Force line breaks with \\
\author{Nishkarsh Kumar}
\affiliation{Department of Physics, Indian Institue of Technology Kanpur, Kalyanpur, India- 208016.}
\author{Jeeban Kumar Nayak}
\email{jkn19rs027@iiserkol.ac.in}
\affiliation{Department of Physical Sciences, Indian Institute of Science Education and Research Kolkata, Mohanpur, India, 741246}
\author{Asima Pradhan}
\email{asima@iitk.ac.in}
\affiliation{Department of Physics, Indian Institue of Technology Kanpur, Kalyanpur, India- 208016.}
\affiliation{Center for Lasers and Photonics (CELP), Indian Institue of Technology Kanpur, Kalyanpur, India- 208016.}
\author{Nirmalya Ghosh}
\affiliation{Department of Physical Sciences, Indian Institute of Science Education and Research Kolkata, Mohanpur, India, 741246}
\begin{abstract} 
Detection of cervical intraepithelial Neoplasia (CIN) at the early stage enables prevention of cervical cancer, which is one of the leading cause of cancer deaths among women. Recently there is a great interest to use the optical polarimetry as a non-invasive diagnosis tool to characterize the cervical tissues. In this context, it is crucial to validate the performance of various Mueller matrix decomposition techniques, that are utilized to extract the intrinsic polarization properties of complex turbid media, such as biological tissues.
%\textbf{Aim}\\
%Quantitative comparison of the extracted polarization parameters from the polar and differential decomposition methods for various optical media. In addition to these quantitative performance of the decomposition techniques, we aim to distinguish between the cervical tissue sections with different grades of CIN by quantifying their polarization properties.\\ 
To quantitatively compare the performance of polar and differential MM decomposition methods for probing the polarization properties in various complex optical media.  Also to utilize the derived individual polarization parameters through differential decomposition method as useful metrics to distinguish between the cervical tissue sections with different grades of CIN.
%\textbf{Approach}\\
Complete polarization responses of the cervical tissue sections, and other media are recorded by preparing a home-built Mueller matrix imaging set up with a spatial resolution of $\approx 400nm$. The Mueller matrices are then processed with the polar and differential decomposition methods to separate, and quantify the individual polarization parameters.
%\textbf{Results}\\
Pronounced differences in the extracted polarization properties are observed for different CIN grades with both the decomposition methods. While a significant increase in the depolarization parameter $(\Delta)$ is observed with progression of CIN stages ($\Delta=0.35$ for CIN-I, and $\Delta=0.56$ for CIN-II), a trend reversal is seen for the linear diattenuation parameter $(d_L)$, where the magnitude of the $d_L$ decreases from $0.26$ to $0.19$ with the growth of CIN, in case of differential decomposition method.\\
%\textbf{Conclusion}\\
Our results indicate that the differential decomposition of MM have certain advantages over the polar decomposition method to extract the intrinsic polarization properties  of a complex tissue medium.  The quantified polarization parameters obtained through the decomposition methods can be used as useful metrics to distinguish between the different grades of CIN, and to describe the healing efficiency of a self-healing organic crystal. Thus the Mueller matrix polarimetry shows great potential as an label-free, non-invasive diagnostic and imaging tool with potential applications in biomedical clinical research, and in various other disciplines.
\end{abstract}
\maketitle
% Include a list of up to six keywords after the abstract
%\keywords{Mueller matrix, Cervical tissue, Tissue polarimetry, Polar decomposition, Differential decomposition}

% Include email contact information for corresponding author
%{\noindent \footnotesize\textbf{*}Jeeban Kumar Nayak,  \linkable{jkn19rs027@iiserkol.ac.in} }
%{\noindent \footnotesize\textbf{+}Asima Pradhan,  \linkable{asima@iitk.ac.in} }
%\begin{spacing}{1}   % use double spacing for rest of manuscript

\section{Introduction}
%\label{sect:intro}  % \label{} allows reference to this section
Many of the natural objects with biological or non-biological origin possess some intrinsic polarization anisotropy such as birefringence, dichroism, and depolarization. Quantification of these polarization anisotropic parameters with desirable spatial resolution, can provide essential morphological structural, and functional information regarding a specimen, and hence quantitative polarimetry has emerged as an powerful tool for the characterization of a wide range of optical media in various disciplines\cite{he2021polarisation,ramella2020review,scatteringreview,gil2022polarized}. In the realm of quantitative polarimetry, polarization resolved Mueller matrix imaging is a widely adopted technique as it enables probing and quantifying complete polarization response of a sample in a single experimental embodiment\cite{he2021polarisation,ramella2020review,ku2019polarization,dengumueller,chipman1995mueller}. Owing to the recent developments in the field of Mueller matrix polarimetry such as, increasing the measurement precision by optimising the calibration methods\cite{he2021polarisation}, adoption of snapshot techniques to extract the polarization information of dynamic objects\cite{hagen2007snapshot}, establishment of various Mueller matrix decomposition and analysis techniques to extract the polarization anisotropy parameters, and etc;\cite{lu1996interpretation}  the MM imaging has gained a special place in biomedical and clinical applications.\cite{he2021polarisation,singh2022mueller}\\
However, carrying out quantitative polarimetry in optically thick turbid media like biological tissues is still a challenge\cite{ghosh2011tissue,novikova2016special,ramella2020review}. Although such complex media possess intrinsic polarization anisotropy, the inevitable depolarization through multiple scattering and simultaneous exhibition of several polarization effects obstruct the quantification of these polarization parameters. Therefore, in recent years a number of Mueller matrix decomposition methods were proposed aiming to separately quantify the polarization anisotropy parameters in a lumped system \cite{polarvsdiff,he2021polarisation,ortega2011mueller,ossikovski2011differential} . These extracted polarization parameters contain a wealth of information regarding the medium properties. The proposed techniques come with different assumptions, and hence one has to be judicious while choosing an appropriate decomposition techniques according to the targeted real-life applications.\\
Some of the recent studies have shown that among the various decomposition methods, the polar and differential matrix decomposition are the efficient techniques to extract the polarization parameters in a complex media, where alongside several polarization anisotropic effects, the medium possess significant depolarization effect. In polar decomposition, a given Mueller matrix is decomposed into sequential products of three basis matrices corresponding to the three general polarization effects (depolarization, diattenuation (linear and circular), retardance (linear and circular)). Yet owing to the non-comuting nature of the matrix multiplication, there is an ambiguity in the derived anisotropic parameters depending on the order of multiplication\cite{chipman1995mueller}. On the other hand differential matrix formalism is a more general kind of decomposition methods, which consider simultaneous exhibition of multiple polarization effects, hence is more appropriate for complex turbid media such as biological tissues. Several researchers have utilized the MM polarimetry for the early stage detection of the cervical cancer.\cite{shukla2009mueller,zaffar2020mapping,zaffar2020assessment}\\

In this work, we have used MM polarimetry to characterize cervical tissue sections with different grades of CIN. Detection of cervical intraepithelial Neoplasia (CIN) at the early stage enables prevention of cervical cancer, which is a leading cause of cancer death among women\cite{cervicaltissue,sung2021global}. Polarization resolved MM imaging of the cervical tissues are recorded, and processed with both polar and differential decomposition methods to extract the associated polarization properties. Individual polarization parameters are obtained by plotting the histogram of spatially quantified polarization images of cervical tissues, and its subsequent fitting with the Gaussian distribution. The obtained parameters are then used as useful metrics to distinguish between the different grades of CIN. Our results not-only enables realization of a non-invasive, label free polarimetric diagnostic tool for the early detection of cervical cancer, but also facilitates quantitative comparison of the performance of various MM decomposition methods in complex turbid media. In addition we have also presented the quantified polarization properties of a self-healing organic crystal to further validate the utilization of different MM decomposition techniques in complex optical media. 
\section{Materials and Methods}
\subsection{Materials}
Sections of $20 \mu m$ thickness from the stromal region of the cervical tissues are used in this study. The cervical tissues are obtained from G.S.V.M. medical college, Kanpur and the sectioning of these cervical tissues are done by Dr. Asha agarwal.
\subsection{Mueller matrix polarimetry set-up}
We have utilized a custom-designed polarization microscopic arrangement, where one can determine the complete polarization response of a sample by recording the $4\times4$ Mueller matrix \cite{sciencemm,acsnano}(Figure 1a). The experimental system employs broadband white light excitation and the subsequent recording of the polarization resolved images of the sample at any selected wavelengths between ($\lambda = 400 - 725 nm$). Alongside imaging, the spectral Mueller matrix of a given sample can also be recorded simultaneously with the Mueller matrix polarimetry system utilized in our experimental configuration. The thirty six projective polarization measurements required for the construction of Mueller matrix is recorded by sequentially generating and analyzing six different linear and circular polarization states\cite{36mm,waveoptics}. A collimated white light from the microscope's (Olympus, IX-71) inbuilt illumination source (halogen lamp, JC 12V 100W) is passed through a polarization state generator unit which comprises of a rotatable (i) linear polarizer and (ii) achromatic quarter wave plate (QWP, Thorlabs AQWP05M-600 ). Polarization of the scattered light is subsequently analyzed by the polarization state analyzer unit that consists of a rotatable achromatic QWP and a linear polarizer. All the optical polarizing elements are mounted on computer controlled motorized rotational mounts (PRM1/M-Z7E, Thorlabs, USA) for precision control. For recording the spectral Mueller matrices of the WPCs, the scattered light is relayed to a spectrometer (HR 4000, Ocean Optics) for spectrally resolved signal detection (resolution $.2nm$). The  Mueller matrix images at selected wavelengths were recorded by using an EM CCD camera (Andor, IXON 3). The polarization resolved intensity images of the cervical tissue sections (dimension = $650\mu m \times 650 \mu m$) are recorded using a $650 nm$ band-pass filter $(\Delta\lambda\approx25nm)$. The imaging Mueller matrices are constructed using 36 polarization-resolved projective measurements (given in the table below)\cite{36mm}.\\
\begin{table*}[ht]
\caption{Scheme for construction of  Mueller matrix using 36 polarization-resolved projective measurements. Here, the first letter represents the input polarization state, the second letter stands for the analyzer or the projected polarization state. The states are defined as $I_H(horizontal)$, $I_V(vertical)$, $I_P (+45\deg)$, $I_M(-45\deg)$, $I_L$ left circular polarized $(LCP)$, $I_R$ right circular polarized $(RCP)$} 
\label{tab:fonts}
\begin{center}       
\begin{tabular}{|c|c|c|c|}  %% this creates two columns
%% |l|l| to left justify each column entry
%% |c|c| to center each column entry
%% use of \rule[]{}{} below opens up each row
\hline
HH+HV+VH+VV & HH+HV$-$VH$-$VV & PH+PV$-$MH$-$MV & RH+RV$-$LH$-$LV\\
\hline
HH$-$HV+VH$-$VV & HH$-$HV$-$VH+VV & PH$-$PV$-$MH+MV & RH$-$RV$-$LH+LV\\
\hline
HP+VP$-$HM$-$VM & HP$-$VP$-$HM+VM & PP$-$PM$-$MP+MM & RP$-$RM$-$LP+LM\\
\hline
HR+VR$-$HL$-$VL & HR$-$VR$-$HL+VL & PR$-$PR$-$MR+ML & RR$-$RL$-$LR+LL \\
\hline
\end{tabular}
\end{center}
\end{table*}

\section{Theory}
Here we will briefly describe the theoretical treatment for the extraction of polarization anisotropy parameters using both polar and differential MM decomposition methods\cite{chipman1995mueller,ghosh2011tissue,polarvsdiff}. In the polar decomposition method, a given Mueller matrix is decomposed into sequential product of three basis matrices.
\begin{equation}
     M = M_\Delta.M_R.M_D
\end{equation}
Where $M_\Delta$ corresponds to the depolarization effect associated with the medium, and $M_R \& M_D$ describe the retardance (linear and circular), and diattenuation (linear and circular) effects respectively. The magnitudes of the anisotropy parameters, diattenuation $(D)$, retardance $(\delta)$, and depolarization $(\Delta)$ are calculated using the respective basis matrices. \\
On the other hand, in the differential matrix formalism, the anisotropic polarization and depolarization effects are stored simultaneously in various elements of a differential matrix $m$ \cite{ortega2011mueller, ossikovski2011differential}. The differential matrix is related to the Mueller matrix $M$ and its spatial derivative along the direction $(\Vec{z})$ of propagation of light as\cite{jones1948new}: 
\begin{equation}
     \frac{dM}{dz}=mM
\end{equation}
The above equation assumes that the sample is laterally homogeneous and both polarization and depolarization effects are occurring simultaneously. In uniformly distributed polarization properties along the propagation direction the integration of differential matrix equation yields:
\begin{equation}
     L=\ln M=m.l
\end{equation}
Here $L$ is the matrix logarithm of recorded Mueller matrix $M$ and $l$ represents the optical pathlength in the medium. The polarization properties of the underlying system can be constructed using the Lorentz symmetric $(L_u)$ and Lorentz antisymmetric $(L_m)$ components of logarithmic Mueller matrix $L=L_m+L_u$ as:
\begin{equation}
\begin{split}
       L_m=1/2(L-GL^TG)\\
       L_u=1/2(L+GL^TG)
\end{split}
\end{equation}
$G$ is the Minkowski metric tensor represented as $G=diag(1,-1,-1,-1 )$. The corresponding anisotropic parameters can be calculated directly from the respective matrix elements of the $L_m$ matrix.\\
\section{Results and Discussion}
The cervix is composed of squamous epithelium, connective tissues, and other components. Collagen is a primary components of the cervical connective tissue, and cross linking between the individual collagen molecules lead to the formation of microfibril and collagen fibers. Anisotropic organization of these collagen fibers makes the cervix polarization anisotropy. Thus, through quantitative MM polarimetric imaging, essential morphological information regarding the cervical tissue sections can be captured. More importantly in case of most cancers, the structural organization of the collagen fibers gets distorted. Consequently it leads to the alteration of the polarization properties of the cervical tissue sections. This enables quantitative MM imaging to be a non-invasive diagnostic tool for pre cancer detection, where quantitative polarimetric analysis can distinguish between a healthy and defected cervical tissue section. \\
Here we have quantitatively analysed the polarization properties of an essential cervical pathologies, the cervical intraepithelial Neoplasia (CIN), which is of particular interest for the early detection of cervical cancer. Polarization MM imaging of cervical tissue sections with CIN grade one (CIN-I) and grade two (CIN-II) are recorded using the MM imaging set up (Fig. 1a). Polarization resolved images of both these tissue sections are captured with a spatial resolution of $\approx400nm$. Consequently the full $4\times4$ MM is constructed from the projective polarization measurements as described in the table 1. The experimentally observed MM for both CIN-I (Fig. 1b) and CIN-II (Fig. 1c) tissue sections demonstrate simultaneous presence of multiple polarization effects such as linear diattenuation $(d_L)$, linear retardance $(\delta_L)$, and depolarization $(\Delta)$. This is evident from the non-zero magnitude of the respective MM elements. Besides the linear diattenuation $(d_L)$ descriptor elements $(M_{12/21}, M_{13/31})$, pronounced magnitude of the elements $(M_{23/32}, M_{24/42})$ describing the linear birefringence $(\delta_L)$ properties are also observed. Significant depolarization of the incident polarized light is also observed, which is expected owing to the multiple scattering in the turbid tissue media.\\
\begin{figure*}
\begin{center}
\begin{tabular}{c}
\includegraphics[height=12cm]{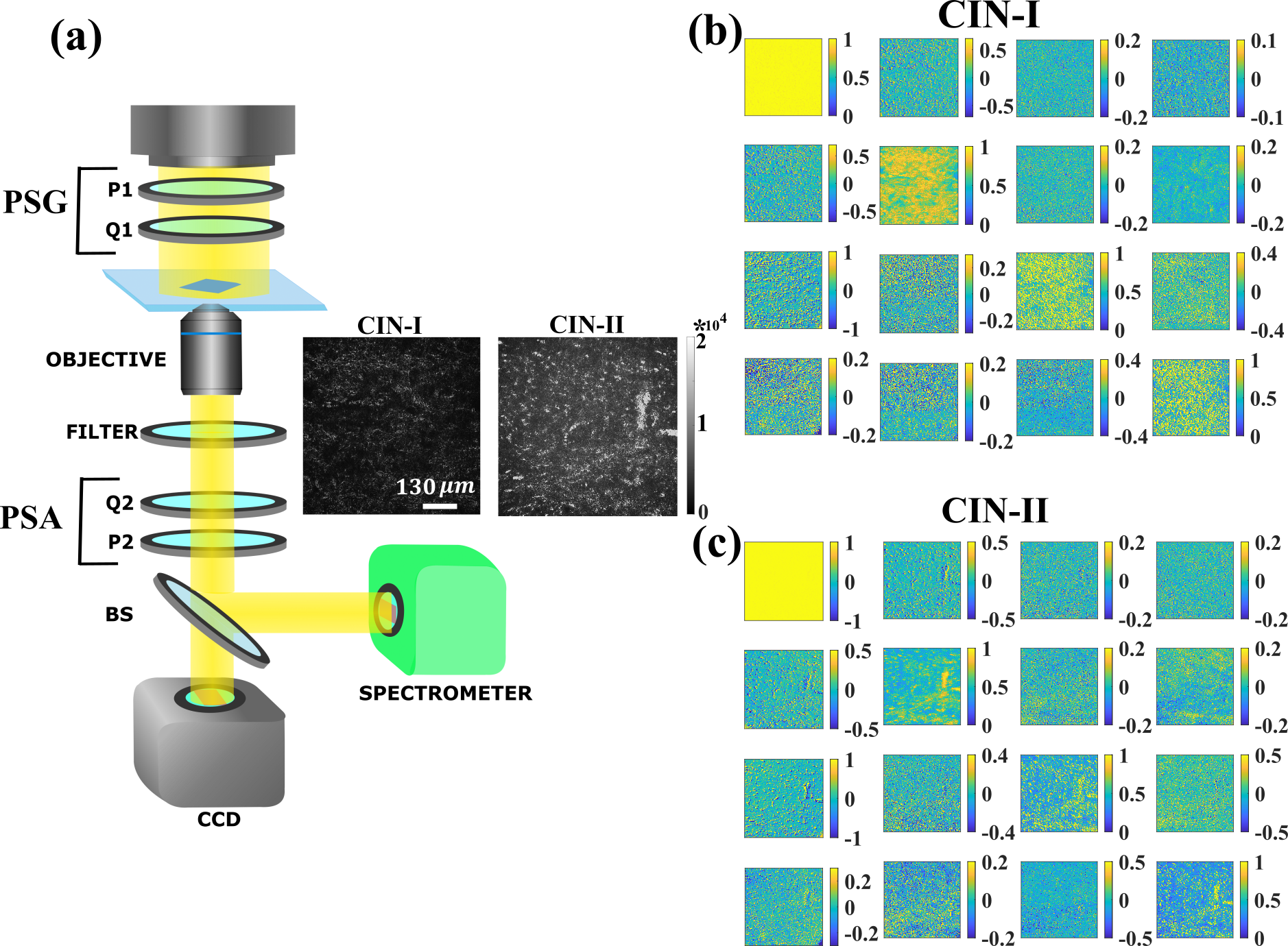}
\end{tabular}
\end{center}
\caption 
{ \label{fig:1}
(a) Schematic of the polarization Mueller matrix microscopy system to capture the spectral and imaging polarization response of a sample. PSG and PSA: Polarization state generator and analyzer units, P: Polarizer, Q: quarter wave-plate, BS: beam splitter. The $4\times4$ polarization Mueller matrix images of the cervical tissue section with CIN grade one (CIN-I) (b), and CIN grade two (CIN-II) (c) are presented. Besides significant depolarization, simultaneous occurrence of multiple polarization effects are manifested in the different Mueller matrix elements. The drawn scale bar is $130 \mu m$.} 
\end{figure*} 
It is pertinent to note that although the intrinsic polarization properties of the cervical tissue sections are reflected in the respective MM elements, the obtained Mueller matrices (fig. 1b, 1c) does not possess a proper symmetry as generally observed for the conventional polarizing optical elements\cite{waveoptics}. This deviation originates due to the presence of significant depolarization effect and simultaneous occurrence of multiple polarization effects, which are manifested in a complex inter-related way in the MM. Thus to extract and quantify the polarization properties, Processing of the observed MM with decomposition methods is indispensable. The derived polarization parameters then can be utilized as useful diagnostic metrics to distinguish between the different grades of CIN, This highlights the significance of MM decomposition technique to establish the quantitative MM polarimetry as a diagnostic tool. The associated decomposed matrices are discussed in the supplementary information. \\
The experimental recorded Mueller matrices (Fig. 1b, 1c) are processed with both polar and differential decomposition methods to obtain the spatial distribution of the individual polarization parameters throughout the tissue sections. Utilizing the corresponding algorithms spatial various polarization anisotropic properties are derived, and the corresponding results are presented in Figure 2. All the obtained polarization parameters; linear diattenuation $(d_L)$ (Fig. 2a), linear retardance $(\delta_L)$ (Fig. 2b), and depolarization $(\Delta)$ (Fig. 2c) show significant strength to be considered as potential metrics to describe the morphological properties of the cervical tissue sections. However, there is an spatial inhomogenity for all the extracted polarization parameters (Fig. 2a(I),(II), b(I)(II), c(I)(II)). Hence we go on to plot  histograms for the respective polarization images, which enables a better quantitative presentation of the polarization properties. Histograms shown in Fig. 2a(III), 2b(III) and 2c(III) are plotted by taking the magnitude of the polarization parameters with no. of pixels attaining that particular magnitude. Furthermore, the histograms are fitted with the Gaussian distribution, and the statistical moments of the distributions (mean $\mu$, standard deviation $(\sigma)$) are calculated. These can serve as  more precise and convenient metrics to distinguish between the different grades of CIN.Comparing the magnitude of the mean value obtained with polar and differential decomposition methods, a significant difference in the magnitude of the linear diattenuation parameter $(d_L)$ is clearly visible in the gaussian fit of the histogram (Fig. 2a(III)). Magnitude of the mean values of $d_L$ correspond to the differential and polar decomposition are found out to be 0.26 and 0.10 respectively. On the other hand, the phase anisotropy parameter $(\delta_l)$  (mean values: $0.13$ and $0.13$) and depolarization (mean values: $0.35$ and $0.32$) does not show much of the variation in the magnitudes for the polar and differential decomposition methods.\\
\begin{figure*}
\begin{center}
\begin{tabular}{c}
\includegraphics[height=12cm]{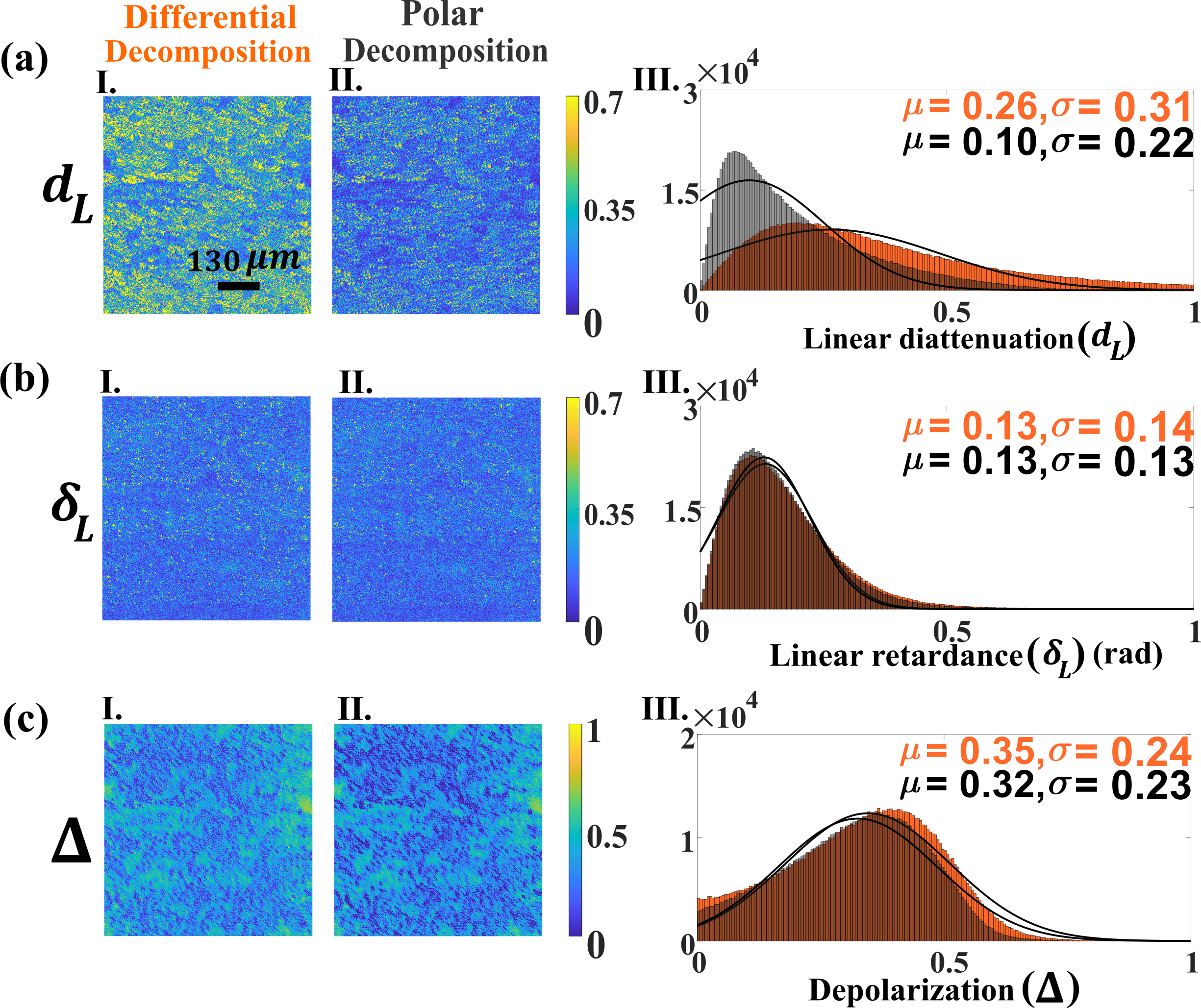}
\end{tabular}
\end{center}
\caption 
{ \label{fig:2}
Intrinsic polarization parameters, such as linear diattenuation $(d_L)$ (a), linear retardance $(\delta_L)$ (b), and depolarization $(\Delta)$ (c) are extracted and quantified from the recorded Mueller matrix of the cervical tissue section of grade CIN-I. While the results obtained for the differential matrix decomposition method is presented in the first column(a(I), b(I), c(I)), the spatial variation of the polarization parameters obtained with the polar decomposition method are given in the second column (a(II), b(II), c(II)). The histogram plots corresponding to the polarization images are presented (a(III),b(III), c(III)), and further the histograms are fitted with Gaussian distribution to calculate the statistical averages (mean $(\mu)$, and standard deviation $(\sigma)$), which are used as metrics to distinguish between the cervical tissue sections. The depicted scale bar is $130 \mu m$} 
\end{figure*} 
The mean values obtained from the Gaussian fitting of the histograms are used to distinguish between the cervical tissue sections with CIN grade one (CIN-I) and two (CIN-II). The results are presented for both polar (Fig. 3a) and differential decomposition method (Fig. 3b), so that besides demonstrating a polarimetric diagnosis tool, we can quantitatively compare the performance of the polar and differential decomposition of MM in case of the cervical tissue sections or in more general for optical interaction in complex turbid media. While the histograms with grey colour corresponds to the CIN-I cervical tissue sections, the magenta colour histograms describe the polarization properties of CIN-II cervical tissue sections. With polar decomposition method, the mean value of the linear diattenuation parameter $(d_L)$ obtained for both CIN-I $(\mu=0.10)$ and CIN-II $(\mu=0.09)$ are nearly equal (Fig. 3a (I)). However, a pronounced difference in the depolarization $(\Delta)$ (Fig. 3a (II))and linear retaradance $(\delta_L)$ (Fig. 3a (III)) parameters between the CIN-I and CIN-II is observed. The mean value of the $\Delta$ significantly increases for CIN-II $(\Delta= 0.53)$ as compared to the CIN-I $(\Delta= 0.32)$ indicating the change in the morphology of the cervical tissue sections. The disruption or randomization of the collagen fibers can lead to the rise of depolarization effect in the cervical tissue. The mean value of the $\delta_L$ also exhibit dissimilarities, where the CIN-II $(\delta_L= 0.21)$ has a relatively higher magnitudes with respect to the CIN-I $(\delta_L= 0.13)$ tissue. Now coming to the MM processing with differential decomposition method, unlike to the case of polar decomposition, there is a pronounced change in the linear diattenuation parameter $(d_L)$ between the CIN-I and CIN-II tissues (Fig. 3b (I)). The obtained mean value of $d_L$ decreases with the progress of CIN stages, and the obtained mean values for CIN-I and CIN-II are $0.26$, and $0.19$ respectively. This describes the degradation of the structural orientation of the anisotropic collagen fibers. The results obtained for depolarization also validates this, where a significant growth in the depolarization magnitude is observed in the CIN-II $(\Delta_L= 0.56)$  as compared to the CIN-I $(\Delta_L= 0.35)$ tissue sections (Fig. 3b (II)). A deviation in the trend is also observed for the case of linear retardance $(\delta_L)$, (Fig. 3b (III)) which is consistent with the results obtained with the polar decomposition method. Altogether, these exclusively demonstrates that, the quantitative MM imaging with incorporation of appropriate decomposition techniques can serve as an efficient label free diagnostic tool for early detection of cervical cancer. We also want to note that the other parameters of the Gaussian distributions such as skewness, kurtosis can also be used as useful metrics to distinguish between the cervical tissues with different grades of CIN.\cite{zaffar2020spatial}\\
\begin{figure*}
\begin{center}
\begin{tabular}{c}
\includegraphics[height=8cm]{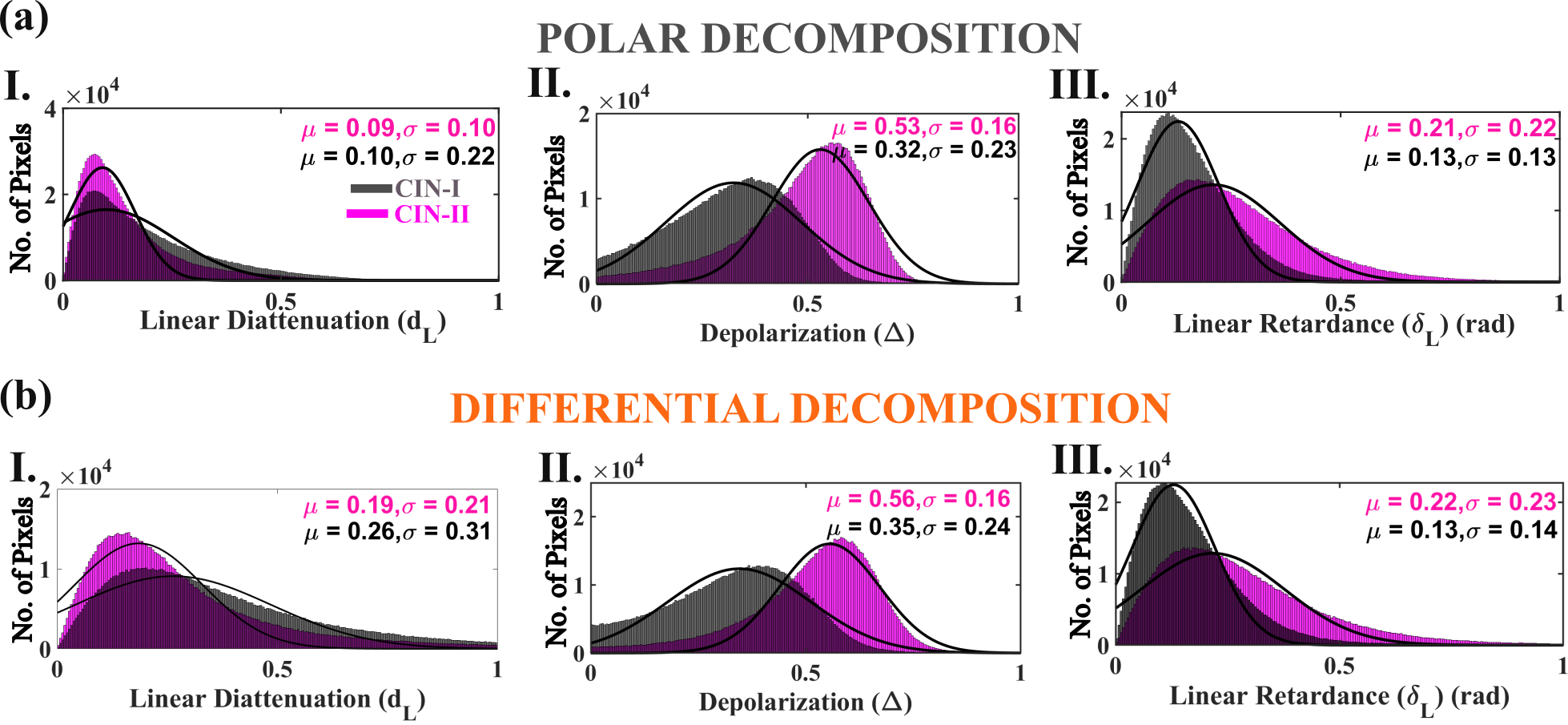}
\end{tabular}
\end{center}
\caption 
{ \label{fig:3}
Histograms corresponding to the spatial variation of the polarization parameters quantified with the polar (a) and differential decomposition method (b) are presented. Comparative evaluation of both the polar and differential decomposition methods are performed by considering the mean values obtained from the Gaussian fitting. Although with polar decomposition, the linear diattenuation $(d_L)$ does not shows much variation for CIN-I and CIN-II, a pronounced change is observed with differential decomposition method (a(I), b(I)). A significant increase in the depolarization parameter with the growth of CIN is observed in the extracted polarization parameter for both polar and differential decomposition techniques (a(II), b(II)). Also increment in the magnitude of the linear retardance $(\delta_L)$ parameter are observed with the progression of CIN stages (a(III), b(III))  } 
\end{figure*} \\
Looking at the results demonstrated above, it is evident that the differential decomposition of MM have certain advantages over the polar decomposition method to extract relevant information regarding the polarization properties of a complex tissue medium. To further compare the performance of both these decomposition techniques, we now consider an anisotropic organic crystal and discuss it's polarization properties.
In our recent work, we have utilized MM polarimetry to probe the healing efficiency of a self-healing bipyrazole piezoelectric crystal. Such extraordinary crystals heal themselves anonymously when subjected to the mechanical fracture through a three-point bending test. Highly ordered crystalline structure of crystals make them a polarization rich entity which shows the strong anisotropic effects. We are examining the different stages of crystals in a similar way as it was shown for the cervical tissues. Three different crystal stages; pristine, neatly healed and imperfectly healed are taken for carrying out the MM imaging, and subsequent quantification of the polarization parameters ($d_L$, $\delta_L$ and $\Delta$) through both polar and differential decomposition methods. In all the scenarios, the magnitude of linear diattenuation $(d_L)$ is found to be very low, showing a very little presence of amplitude anisotropy due to lack of the imaginary part of the refractive index. Hence the other two polarization parameters (linear retardance $(\delta_L),$, and depolarization $(\Delta_L)$) are considered for the comparative study of the crystals in different stages. The corresponding results are shown in figure 4. A region of $26\mu m \times 104 \mu m$ dimension (shown with a rectangle in images of the Figure 4) is selected in each crystal through the microscope eyepiece and verified by calculating the higher correlation values for the specific dimensions between the pristine and healed crystals. Linear retardance mean values with their standard deviation for the pristine and neatly healed crystal were $0.42\pm 0.04$  and $0.34\pm0.05$ respectively resulting in the retrieval $(80 - 85\%)$ of the phase anisotropy of the crystal in healing process (Fig. 4a(II),(IV), 4b(II)(IV)). Depolarization $\Delta$ shows the comparable values for the pristine $ (0.46\pm0.03)$ and healed $ (0.44\pm0.03) $ (Fig. 4a(I)(III), 4b(I)(III)) . In case of the imperfectly healed, crack junction is clearly observed in the crystal (Fig. 4c), $\delta_L$ in left and right part of the crystal is remain intact with respect to the pristine while depolarization magnitude has increased. While being under mechanical fractures, the order of the crystalline structure gets scrambled leading to the decrease in anisotropic parameters and store their order as the crystal heals itself. Orientation angle of linear retarder axis provide a better understanding for the efficient repairing of the fractured crystal (Fig. 4a(V), 4b(V), 4c(V)). While the  long-range crystalline order remains intact for both pristine (Fig. 4a(V)) and the healed crystal (Fig. 4b(V)), a clear disorder in the orientation of the retarder axis is observed for the imperfectly healed crystal (Fig. 4c(V)). Important thing to note here is that, both the polar and differential decomposition of the MM gives rise to the near-equal magnitude of the extracted polarization parameters which was not the case for the cervical tissue sections. A table containing all the mean and standard deviation values of the respective polarization parameters is given in the supplementary information.\\
\begin{figure*}
\begin{center}
\begin{tabular}{c}
\includegraphics[height=10cm]{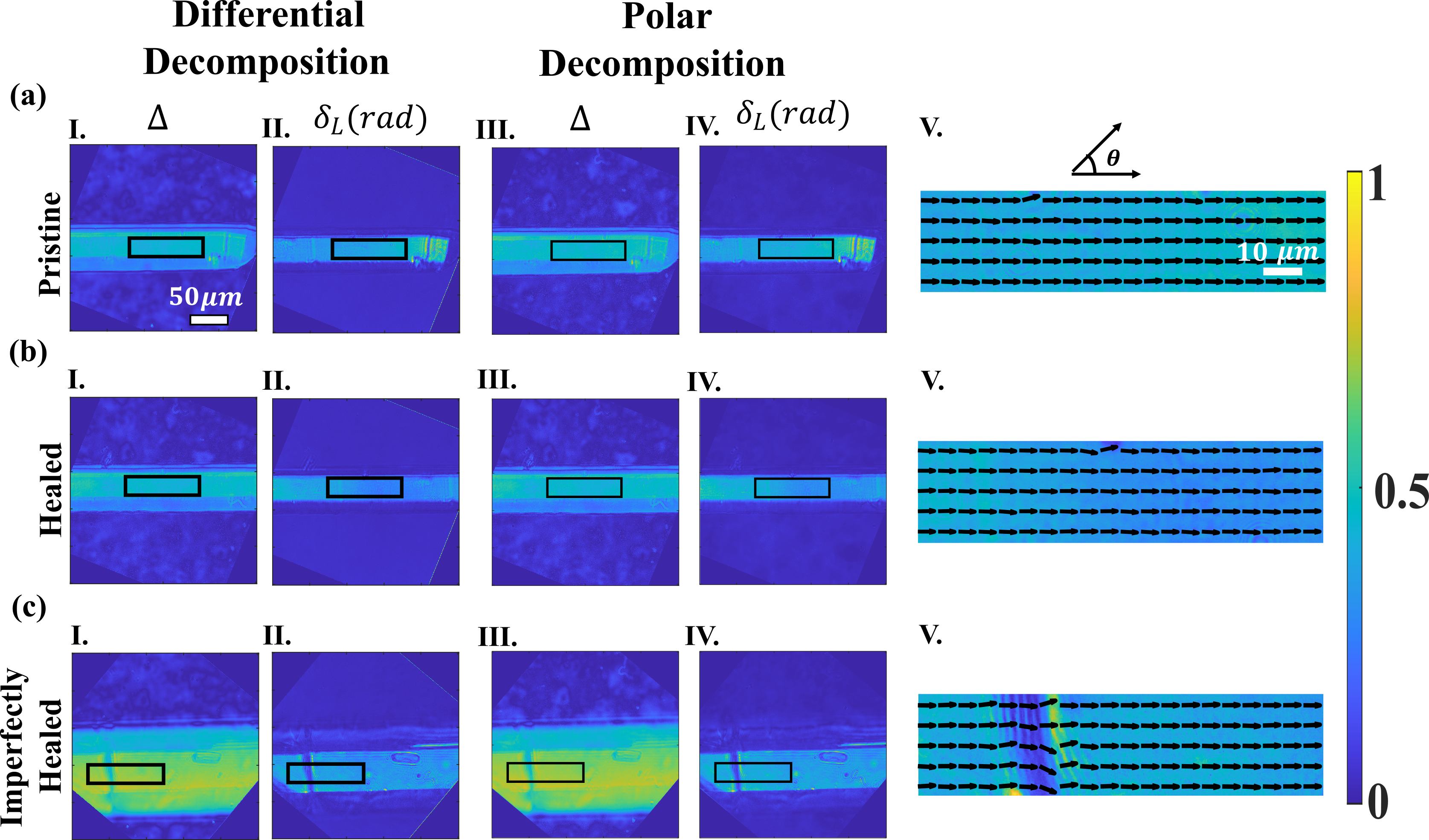}
\end{tabular}
\end{center}
\caption 
{ \label{fig:4} Quantified polarization parameters of the pristine (a), healed (b) and imperfectly healed crystal (c) are provided for both polar and differential decomposition of the MM. Spatial variation of the depolarization $(\Delta)$ (I.)(II.) and linear retardance $(\delta_L)$ (III.)(IV.) parameters  are extracted from the inverse analysis of the Mueller matrix. Variation in such homogeneous organic crystals is negligible while processing through different analysis methods. An area  with dimensions $26\mu m \times 104 \mu m $ (black colored box present in  each image) has been analyzed for the quantitative comparison of the derived polarization parameters between the pristine, healed, and imperfectly-healed crystal. (V) Spatial variation of the orientation axis of the linear retarder is also plotted for the considered region of interest (black solid box), where a clear discontinuity is visible along the crack region (c(V.)) demonstrating the decrease in the linear retardance parameter. Scale bar is $50\mu m$.}
\end{figure*} \\
\section{Conclusion}
In summary, we have demonstrated that the differential decomposition of MM have certain advantages over the polar decomposition method to extract the intrinsic polarization properties of a complex turbid media. For this purpose, cervical tissue sections with different grades of CIN are considered, which alongside significant depolarization, exhibits multiple polarization effect simultaneously. The polarization properties of these cervical tissue sections are quantified through the incorporation of both polar and differential MM decomposition methods. We have shown that, the individual derived polarization parameters through differential decomposition can be treated as useful metrics to distinguish between different stages of CIN, which enables realization of a non-invasive, label-free diagnosis tool for early detection of the cervical cancer. In addition we have also investigated the performance of both polar and differential decomposition MM methods to quantify the healing efficiency of a self-healing organic crystal.  This highlights the ability of MM polarimetry with an appropriate decomposition technique to be an efficient diagnosis and imaging tool with potential applications in various disciplines. 
\newpage
%\subsubsection{Reference linking and DOIs}
%A Digital Object Identifier (DOI) is a unique alphanumeric string assigned to a digital object, such as a journal article or a book chapter, that provides a persistent link to its location on the internet. The use of DOIs allows readers to easily access cited articles. Authors should include the DOI at the end of each reference in brackets if a DOI is available. See examples at the end of this manuscript. A free DOI lookup service is available from CrossRef at \\\linkable{http://www.crossref.org/freeTextQuery/}. The inclusion of DOIs will facilitate reference linking and is highly recommended. 
%In the present LaTeX template, the author needs to add the DOI reference by including it in a ``note'' in the bibliography file, as shown in the file {\verb+report.bib+}, for example, \\ {\verb+note = "[doi:10.1117/12.154577]"+}. The DOI may be used by the reader to locate that document with the link: {\verb+http://dx.doi.org10.1117/12.154577+}. 
%\subsection{Biographies}
%A brief professional biography of approximately 75 words may be provided for each author, if available. Biographies should be placed at the end of the paper, after the references. Personal information such as hobbies or birthplace/birthdate should not be included. Author photographs are not published.
% \disclosures 
\section*{Disclosures}
The authors declare no conflict of interests.
\section* {Code, Data, and Materials Availability} 
The data, utilized algorithms for the decomposition of Mueller matrix supporting the results that have been reported in this study are available from the corresponding authors upon reasonable request.
\section* {Acknowledgments}
Jeeban Kumar Nayak and Nirmalya Ghosh acknowledge the support of the Indian Institute of Science Education and Research (IISER) Kolkata, an autonomous institute under the Ministry of Human Resource Development (MHRD), Government of India. Nishkarsh Kumar and Asima Pradhan acknowledge the support from Indian Institue of Technology (IIT) Kanpur. The authors acknowledge Dr. Kiran Pandey from GSVM medical college for providing the cervical tissues and Dr. Asha Agarwal for the histopathology sectioning. We acknowledge Dr. Chilla Malla Reddy and Dr. Surojit Bhunia for providing the self-healing crystals.
\newpage
%%%%% References %%%%%

\bibliography{report}   % bibliography data in report.bib
%\bibliographystyle{spiejour}   % makes bibtex use spiejour.bst

%%%%% Biographies of authors %%%%%
%\section{Biography}
%\textbf{Nishkarsh Kumar} is pursing his PhD at the University of Optical Engineering. He received his BS and MS degrees in physics from IISER KOLKATA in 2020. His research interests include.\\
%\textbf{Jeeban Kumar Nayak} is a senior research fellow at the BIONAP lab in the department of physcial sciences IISER KOLKATA. He received his BS and MS degrees in physics from IISER KOLKATA in 2019. His current research interests are polarization optics, differential imaging, and spin-orbit photonics in liquid crystal and meta-surfaces.\\
%\textbf{Asima Pradhan} is currently a professor in the . She received her PhD from. Her current research interests are polarization optics, differential imaging, and spin-orbit photonics in liquid crystal and meta-surfaces.\\
%\textbf{Nirmalya Ghosh} is currently a professor in the department of Physical sciences at IISER KOLKATA. 
%\vspace{1ex}
%\noindent Biographies and photographs of the other authors are not available.
%\listoffigures
%\listoftables

%end{spacing}
\end{document}